# Negative Group Velocity from Quadrupole Resonance of Plasmonic Spheres

Dezhuan Han, Yun Lai, Kin Hung Fung, Zhao-Qing Zhang and C. T. Chan

Department of Physics
The Hong Kong University of Science and Technology
Clear Water Bay, Kowloon, Hong Kong, China

## Abstract

We study the dispersions of plasmonic bands that arise from the coupling of electric quadrupole resonances in three dimensional photonic crystals (PCs) consisting of plasmonic spheres. Through analytical derivation, we show that two branches of quadrupole bands in simple cubic PCs with a small lattice constant possess negative group velocities. Distinct from double negative ($\varepsilon$, $\mu$ <0) media in which the negative responses originates from the coupling of electric and magnetic responses (**P** and **M**), the negative dispersion induced by quadrupole resonance is an intrinsic property of quadrupole that does not require coupling to another degree of freedom. In addition, there is no simple effective medium description. In plasmonic systems composed of metallic nanoparticle clusters, the coupled quadrupole resonance may be tuned to lower optical frequencies, and the coupling strength between this quadrupole resonance and external electromagnetic (EM) waves are in the same order of the magnetic dipole **M**.

## I. INTRODUCTION

Negative refraction at a particular frequency range can be achieved if there are simultaneously negative values of electric permittivity $\varepsilon$ and magnetic permeability $\mu$ [1]. Such kind of material has interesting optical properties, including the possibility of transmitting both propagating and evanescent EM waves [2]. It is relatively straightforward to obtain negative permittivity, but it is technically difficult to achieve negative permeability especially at optical frequencies. Recently negative refractive index (negative-$n$) materials are demonstrated from microwave to optical frequencies by introducing resonant elements such as split-ring resonators (SRR) [3], and nano-plasmonic structures such as nanorod pairs [4] and "fishnet" structures [5]. In optical frequencies, the plasmomic clucters such as ring structure [6,7] are also plausible candidates to achieve magnetic resonance. Yet one have to face the difficulty to define a meaningful magnetic permeability $\mu$, as discussed by Landau [8], since the magnetic response in optical frequencies comes from the curl of electric dipoles. In high frequencies, it may be more convenient to consider the spatial dispersion [9] of the dielectric function $\varepsilon(\omega,\mathbf{k})$ which can describe comprehensively the electric and magnetic response. The electric quadrupole $\mathbf{Q}$, generated by the symmetric part of the field gradient, $sym(\nabla \mathbf{E})$, can be treated as the lowest order effect which yields spatial dispersion. Another point should be emphasized is that, in plasmonic structures such as clusters (ring structure, etc.), $\mathbf{Q}$ is always in the same order of magnetic moment $\mathbf{M}$. In the parallel metallic nanobar structure, the contribution from $\mathbf{Q}$ has also been proved to be comparable to that from $\mathbf{M}$ in the far field [10].

In plasmonic lattices, the dispersion can be viewed as the consequence of the hopping of resonances from one unit to another, and in that sense, plamonic systems are analogous to electrons, where the band structure originates from the hopping of electrons from one site to another in the tight-binding picture. We may thus associate the plasmonic dispersions that are derived from dipoles and quadrupoles with electronic bands of $p$ and $d$ states respectively. In electronic systems, $sp$-states are dispersive and have wide band widths typically, while the $d$-states are more localized leading to smaller band width [11]. Similar behaviors are expected in plasmonic systems so that the dipole bands should have wider band width than the quadrupole bands. In some transition metals, there are strong hybridization between $sp$ states and the *d* states. In plasmonics, the hybridization can be tuned as will since the lattices are artificial.

In this article, we will investigate the dispersion properties of three-dimensional plasmonic lattices made up of small metal nanospheres. Such a problem is typically considered by treating each nanosphere as a dipole. We will give a thorough analysis by including the quadrupoles. It is shown that the quadrupole resonance in simple cubic PCs can lead to a new kind of negative band which is different from the Veselago type with double negative values of $\varepsilon$ and $\mu$, or the Bragg scattering type [12,13] since here the lattice constant *a* is much smaller than wavelength $\lambda$.

This paper is organized as follows. In Sec. II, we calculate the band structures of three-dimensional simple cubic PCs. Bands with negative group velocity are observed independent of the lattice constant *a*, and these negative bands are found to originate from quadrupole resonances. In Sec. III, analytical modeling is developed which is analogous to the coupled dipole equations [14,15]. The negative group velocities $v_g$ are derived analytically, and the anisotropy of the bands are found even in the long wavelength limit. In Sec. IV, We further considered the plasmonic quadrupole mode dispersion of a nanoparticle cluster, using an octahedral cluster as prototype. In Sec. V, we conclude this work and in appendixes we give mathematics details omitted from main text.

## II. QUADRUPOLE BANDS

We first calculate the band structures of simple cubic PCs by using the multiple scattering theory (MST) [16]. We note that MST for spherical objects is based on the expansion of the fields in terms of vector spherical harmonics, and the convergence is controlled by the number of angular momentum channels used. If the calculation is truncated so that only spherical harmonics with *l*=1 are used, the results will be equivalent to using a dipole approximation, while if only *l*=2 terms are used, it is a quadrupole approximation. It is known in the literature that for metallic spheres arranged in one dimensional chain, the dipole and quadruple dispersions are separated in energy if $r_s \leq a/3$ [17]. In this article, we use Drude type permittivity $\varepsilon(\omega) = 1 - \omega_p^2/\omega^2$ to describe the response of the metal with a bulk plasma frequency of $\omega_p$ = 6.18 eV. The salient features of the results remain the same for other choices of the plasma frequency. In Fig. 1, we show the band structures for two values of the lattice constant *a* (20 nm and 60 nm) while keeping the ratio of the radius to lattice constant $r_s/a = 0.26$ fixed. We find a total of eight bands, three of them are of dipole origin, and the other five are quadrupole bands. Counting the bands from low energy to higher energy, the two low-lying bands are transverse dipole bands, in which the dipoles are perpendicular to the k-vectors. We then have the longitudinal dipole band at higher energy, with the dipoles pointing along the k-vectors. The five quadrupole bands near 3.9 eV are less dispersive. Bands with negative group velocity near the zone center are observed for two branches of the quadrupole bands. The calculations show that the band structures with *l*=1,2 have only very slight modifications if higher multipoles $l \geq 3$ are included. This is not surprising if we note that the quadrupole bands do not reach the octupole (*l*=3) resonance frequency $\omega_{l=3} \sim \sqrt{\frac{l}{2l+1}}\omega_p = 4.05$ eV. The quadruple bands are found to be essentially scale invariant, in the sense that the dispersion is nearly the same in an absolute frequency scale if we vary the lattice constant but keep the ratio $r_s/a$ to be the same. However, a small redshift about 0.02 eV in photon energy are observed in Fig. 1 when sphere radius increases from 5.2 to 15.6 nm. This redshift comes from the redshift of single sphere resonance frequency when the sphere radius becomes larger [14]. The longitudinal dipole mode is also very robust, while the transverse modes change more conspicuously because they are coupled to free photons whose dispersion is given by the light line, which move closer and closer to vertical axis with the slope proportional to $\pi/a$ as the lattice constant becomes smaller. As a consequence, the dispersion of the transverse dipole modes have larger slope when *a* becomes smaller as shown in Fig. 1. Similar phenomena can be observed for longitudinal and quadrupole modes near 3.8 and 3.9 eV as light line intersects the plasmonic resonances.

   Let us now consider the hybridization of the **P** and **Q** modes. The polarizabilty $\alpha_P$, $\alpha_Q$ for **P**, **Q** are proportional to the corresponding Mie coefficients [18] such that $\alpha_P(\omega) = 3ia_1(\omega)/2k_0^3$, and $\alpha_Q(\omega) = 30ia_2(\omega)/k_0^5$, where $k_0 = \omega/c$ is the wavevector in vacuum. By setting $\alpha_{P(Q)} = 0$, we can obtain "pure quadrupole" (or "pure dipole") dispersion respectively. Fig. 2 compares the complete band structure (black), which treats all angular momentum channels on equal footing, with pure dipole (green) and pure quadrupole (red) bands. We can see that for the same lattice constant, the small radius [Fig. 2(a)] configuration shows almost no hydridization between the **P** and **Q** bands, while some hydridization is observed if the sphere has a larger radius [Fig. 2(b)]. More information can be obtained by the eigenstate analysis. Let us focus on the quadrupole bands. In Fig. 2(a), taking the ΓX direction as example, the lowest energy band is a $Q_{yz}$ quadrupole band with a small negative $v_g$. At higher energies the bands correspond to the two-fold degenerate states $Q_{xz}$ and $Q_{xy}$. The highest energy bands are the $Q_{yy} = \pm Q_{zz}$ bands with negative $v_g$. We note that $Q_{xx}$ is not a linear independent variable since $\text{tr}\mathbf{Q} = 0$. Three off-diagonal **Q** states are degenerate at the Γ point and so are the two diagonal **Q** states, as required by symmetry. In Fig. 2(a), the agreement of black and green (red) dots indicates the independence of **P** and **Q** bands when the ratio $r_s/a$ is 0.26. The hybridization between **P** and **Q** can be neglected. As $r_s/a$ increases to 0.3, longitudinal **P** and $Q_{yy} = Q_{zz}$ mode intersect at about 3.97 eV, which is shown in Fig. 2(b). The anti-crossing effect can be observed and the original **P**, **Q** bands hybridize to form two new bands. In order to make the physics simple, we focus the following discussions on configurations which has a clean separation of **P** and **Q** derived bands.

### III. ANALYTICAL MODELLING

To obtain a better understanding of the band dispersion due to the quadrupoles, we will set up a set of coupled quadruple equations. This is essentially the same strategy of setting up coupled dipole equations to study the dispersion of dipolar arrays. We start from the electric fields at point **r** generated by the quadrupole at the origin:

$$\mathbf{E}(\mathbf{r}) = -\frac{i}{6}\left[\left(f'(r) - 2f(r)/r\right)(\mathbf{n}\cdot\mathbf{Q}\cdot\mathbf{n})\mathbf{n} - \left(f'(r) + f(r)/r\right)\mathbf{Q}\cdot\mathbf{n}\right], \quad (1)$$

where $f(r) = k_0^3 h_2^{(1)}(k_0 r)$, $\mathbf{n} = \mathbf{r}/r$ is an unit vector, and $\alpha, \beta$ are indices of coordinates which run from 1 to 3. We use the function $f(r)$ with argument $r$ instead of spherical Hankel function $h_2^{(1)}(k_0 r)$ with argument $k_0 r$ for the sake of convenience. Since the quadrupoles respond linearly to the symmetric part of the gradient of the local electric field, we define a second-rank tensor field **F** as

$$F_{\alpha\beta} = sym(\nabla \mathbf{E}) = \frac{1}{2}\left(\partial E_\alpha/\partial x_\beta + \partial E_\beta/\partial x_\alpha\right). \quad (2)$$

The gradient of field $\nabla \mathbf{E}$ can be calculated from Eq. (1). The diagonal and off-diagonal terms of **F** are respectively

$$F_{\alpha\alpha} = -\frac{i}{6}\left[(f_1 n_\alpha^2 + f_2)\mathbf{n}\cdot\mathbf{Q}\cdot\mathbf{n} + f_3 n_\alpha(\mathbf{Q}\cdot\mathbf{n})_\alpha + f_4\cdot Q_{\alpha\alpha}\right]$$
$$F_{\alpha\beta} = -\frac{i}{6}\left[f_1 n_\alpha n_\beta \mathbf{n}\cdot\mathbf{Q}\cdot\mathbf{n} + \frac{1}{2}f_3\left(n_\beta(\mathbf{Q}\cdot\mathbf{n})_\alpha + n_\alpha(\mathbf{Q}\cdot\mathbf{n})_\beta\right) + f_4 Q_{\alpha\beta}\right], \quad (3)$$

where $\nabla\cdot\mathbf{E}=0$ will be satisfied automatically since $\text{tr}\mathbf{Q}=0$, and no Einstein summation convention is assumed. In Eq. (3), we define $f_1$, $f_2$, $f_3$, $f_4$ as auxiliary functions which are linear combinations of $f(r)$, $f'(r)$ and $f''(r)$:

$$f_1 = f''(r) - 5f'(r)/r + 8f(r)/r^2, \quad f_2 = f'(r)/r - 2f(r)/r^2,$$
$$f_3 = -f''(r) + 2f'(r)/r - 2f(r)/r^2, \quad f_4 = -f'(r)/r - f(r)/r^2.$$

From Eq. (3) the Green function $\mathbf{G}^Q(\mathbf{r}\text{-}\mathbf{r}')$ for the second-rank tensor field **F** can be defined, which satisfies

$$\mathbf{F}(\mathbf{r}) = \mathbf{G}^Q(\mathbf{r}\text{-}\mathbf{r}')\,\mathbf{Q}(\mathbf{r}'). \quad (4)$$

To deal with the second rank tensor in a compact form, we can rewrite **Q** and **F** as vectors: $\mathbf{Q} = (Q_{22}, Q_{33}, Q_{12}, Q_{13}, Q_{23})^T$ and $\mathbf{F} = (F_{22}, F_{33}, F_{12}, F_{13}, F_{23})^T$. In this vector notation, the Green function $\mathbf{G}^Q(\mathbf{r}\text{-}\mathbf{r}')$ is reduced to a 5×5 matrix.

Once the local tensor field **F** is given, the linear response is simply $\mathbf{Q} = \alpha_Q \mathbf{F}$. Considering a quadrupole at position $\mathbf{r}_i$ in the lattice, the corresponding local tensor field **F** is generated by all the other quadrupoles except itself, namely

$$\mathbf{Q}(\mathbf{r}_i) = \alpha_Q \mathbf{F}(\mathbf{r}_i) = \alpha_Q \sum_{\mathbf{r}_j \neq \mathbf{r}_i} \mathbf{G}^Q(\mathbf{r}_i - \mathbf{r}_j)\mathbf{Q}(\mathbf{r}_j). \quad (5)$$

The Green function $\mathbf{G}^Q(\mathbf{r}\text{-}\mathbf{r}')$ can be obtained from Eq. (3). With the Bloch condition imposed on Eq. (5), we will reproduce the rigorous results given by three dimensional MST with $l=2$ only. (Since $\mu=1$ for the plasmonic spheres here, the contribution from magnetic quadrupole is not considered here.) We see from Fig. 2(a) that for plasmonic lattices in which the $r_s/a$ ratio is smaller than 0.26, the bands for $l=1$ and $l=2$ has no observable hybridization, and the two angular momentum channels are indeed separable. We can thus discuss "quadruple bands" and "dipole bands" separately. For the quadrupole bands, the dispersion is analogous to *d*-bands in transition metals [11]. For higher angular momentum states, the coupling is typically short-ranged. The short range interaction is dominated by the near zone field in electrodynamics, or in other words, the quasi-static limit [19,20] of EM fields. We will show that the quadrupole bands obtained within the framework of the quasi-static approximation (QSA) agree very well with the full electrodynamic results.

In the quasi-static limit, we take the light speed $c$ to infinity and the electric field **E** does not depend on $k_0$. In that limit, only the terms containing the lowest power of $r$ survives, and the auxiliary functions $f_i$ are reduced to the simple forms

$$f_1 \sim -105i/r^5,\ f_2 \sim 15i/r^5,\ f_3 \sim 60i/r^5,\ f_4 \sim -6i/r^5. \quad (6)$$

In the quasi-static limit, all slowly convergent powers of $r$ in $\nabla \mathbf{E}$ drop out, and we need to consider only $r^{-5}$ terms, which simplify the lattice sum in Eq. (5) considerably. Special summation techniques developed to accelerate convergence such as Ewald's sum is unnecessary in the QSA here. Eqs. (3) and (6) give us the quadrupole's Green function in QSA with a simple 5×5 matrix form. As is known from the band structure analysis above, the eigenstates of the two quadrupole negative bands contain only the diagonal elements $\{Q_{yy}, Q_{zz}\}$ of the **Q** tensor. The subspace spanned by their eigenvectors is orthogonal to that formed by the non-diagonal elements. The bands of these $\{Q_{yy}, Q_{zz}\}$ states are of special interest since they have intrinsic negative group velocities. **F**, **Q** are reduced to two dimensional vectors when we focus on the subspace spanned by $Q_{yy}$ and $Q_{zz}$ that give two negative group velocity quadruple bands. We define two dimensional vectors $\mathbf{Q}'=(Q_{yy},Q_{zz})$, $\mathbf{F}'=(F_{yy}, F_{zz})$, and the 2×2 matrix form of QSA Green function becomes

$$\mathbf{G}'(\mathbf{r}) = \frac{1}{6r^5} \begin{bmatrix} 105n_y^2(n_x^2-n_y^2) + 75n_y^2 - 15n_x^2 - 6 & 15(n_x^2-n_z^2)(7n_y^2-1) \\ 15(n_x^2-n_y^2)(7n_z^2-1) & 105n_z^2(n_x^2-n_z^2) + 75n_z^2 - 15n_x^2 - 6 \end{bmatrix}, \quad (7)$$

where **n** is again the unit vector in the direction of **r**.

To find the Bloch states, the lattice Green function is transformed to the Fourier space: $\mathbf{G}'(\mathbf{k}) = \sum_{\mathbf{r}_i \neq 0} \mathbf{G}'(\mathbf{r}_i)\, e^{i\mathbf{k}\cdot\mathbf{r}_i}$. Eq. (5) becomes a secular equation of the form $\det\left(\alpha_Q^{-1}\mathbf{I} - \mathbf{G}'(\mathbf{k})\right) = 0$, where $\alpha_Q^{-1}$ and $\mathbf{G}'(\mathbf{k})$ contain the material and geometric information respectively. The dyadic Green function in (7) now has a simple structure since the only dependence on the lattice constant is the prefactor $r^{-5}$. We can evaluate the lattice sum in a mesh of integral points $(i, j, k)$ independent of the lattice constant to solve the secular equation. In Fig. 3(a), we can see that the QSA dispersion relations (green dots) for the states spanned by $Q_{yy}$ and $Q_{zz}$ agree well with the MST results (black dots). Since the QSA gives such a good agreement for the quadrupole bands, we can now discuss the physics using the QSA picture to give a more straightforward interpretation and we can give analytic results near the zone center. At the $\Gamma$ point we can evaluate the lattice Green function semi-analytically, which is $\mathbf{G}'=-a^{-5}\sigma\cdot\mathbf{I}$, and $\sigma$ is the lattice sum:

$$\sigma = \frac{1}{6(i^2+j^2+k^2)^{5/2}} \sum_{(i,j,k)\neq 0} \frac{105 j^2(i^2-j^2)/(i^2+j^2+k^2)^2}{+(75j^2-15i^2)/(i^2+j^2+k^2)-6},$$

which is easily summed to be 18.65. This is a constant due to the geometry of the system. The quadrupole polarizabilty in QSA has a simple form $\alpha_Q^{-1} \to (1-5\omega^2/2\omega_p^2)r_s^{-5}$ if the Drude-type permittivity is applied. Combining the dynamical and geometric parts in the secular equation, we get the QSA frequency of eigenstate at the $\Gamma$ point

$$\omega_0 = \omega_p\sqrt{\frac{2}{5}\left(1+\sigma\left(r_s/a\right)^5\right)}. \qquad (8)$$

The term of $\left(r_s/a\right)^5$ can be regarded as the geometric correction of the single sphere quadrupole resonance $\omega_{l=2} = \sqrt{\frac{2}{5}}\omega_p$. This semi-analytic result indicates the symmetry required degeneracy of two eigenstates $Q_{yy}=\pm Q_{zz}$ at the $\Gamma$ point for the simple cubic lattice, which is manifested in the numerical calculations, such as the example shown in Fig. 1 with $r_s/a = 0.26$. The result given by formula (8) $\omega_0$ =3.95 eV is the same as the numerical result from MST up to two decimal places when $a$ = 20 nm.

To show that the group velocity is intrinsically negative, we find the group velocity and the second derivative of the photon dispersion near the $\Gamma$ point. We do the first and second derivatives on the secular equation. The first derivative $\nabla_\mathbf{k}\mathbf{G}'$ is found to be exactly zero, thus the group velocity is $\partial\omega/\partial k = 0$ as expected. The second order derivative of the dispersions at the zone center are found to be direction dependent. Taking $\Gamma X$ and $\Gamma M$ direction as examples, we obtain $\partial^2\mathbf{G}'/\partial k^2 \big|_{\Gamma X} = a^{-3}\begin{bmatrix} 4.71 & 2.86 \\ 2.86 & 4.71 \end{bmatrix}$, and $\partial^2\mathbf{G}'/\partial k^2 \big|_{\Gamma M} = a^{-3}\begin{bmatrix} 4.71 & -1.43 \\ -1.43 & 4.71 \end{bmatrix}$. The second derivative of secular equation finally give us the results:

$\partial^2\omega/\partial k^2|_{\Gamma X} = (-4.71 \pm 2.86)m_c^{-1}$ and $\partial^2\omega/\partial k^2|_{\Gamma M} = (-4.71 \pm 1.43)m_c^{-1}$, where $m_c = 5\omega_0 a^3/\omega_p^2 r_s^5$ is a system parameter. This shows that the dispersion is necessarily anisotropic, even for a system with cubic symmetry. The inverse of second derivative of the plasmonic dispersion corresponds to the effective mass in electronics. The band with larger absolute value of second derivative has more negative slope and lower frequency. Also it is clear that the contrast $m_{\Gamma X}/m_{\Gamma M}$ is closer to 1 for the lower frequency band. The equifrequency contours of these negative bands cannot be simply represented by ellipsoids. In PCs, similar anisotropy near the $\Gamma$ point have also been be observed [21], which is essentially introduced by multipoles with $l \geq 2$. Fig. 3(c) shows the equifrequency contours of the two negative bands. The lower frequency band with smaller negative slope has more circular shape as expected, while the higher frequency band is distinctly anisotropic. By checking the eigenstates, the lower frequency band corresponds to the eigenstates $Q_{yy} = Q_{zz}$ in the $\Gamma X$ direction and $Q_{yy} = -Q_{zz}$ in the $\Gamma M$ direction respectively.

The above results show that the bands derived from $Q_{yy}$ and $Q_{zz}$ are intrinsically negative in group velocity in a simple cubic array. We should emphasize here that the negative group velocity bands differ fundamentally from that given by other mechanisms. For example, we can obtain negative group velocity in a lattice by Bragg scattering which is typically the case in dielectric PCs. In those cases, the frequency of negative dispersion bands must be scale dependent, which means that the frequency scales with the lattice parameter. However, for plasmonic lattices, the dispersions are nearly scale independent, as shown in Fig. 1. Another way to obtain negative dispersion is through "double negativity" by the coupling of electric and magnetic dipoles, as proposed by Veselago [1]. In that case, the electric dipole and magnetic dipole resonances are strong enough that they each give respectively negative $\varepsilon$ and $\mu$. Although each resonance **P** or **M** on its own gives a band gap, when they couple together in the same frequency, a pass band with negative group velocity appears. The intrinsic negative group velocity bands due to the electric quadrupoles are different since the mechanism is not the coupling of electric and magnetic resonances, but the hopping of one type of electric resonance (quadrupole) from one site to another. In this aspect, the physics is rather similar to tight-binding picture of electrons. We note in particular that it is not meaningful to extract effective permittivity or permeability from such as system. If we insist on applying inversion procedures (say by using the amplitude and phase of transmission and reflection to extract some form of negative $\varepsilon$ and $\mu$ [22], field averaging method [23], or other effective medium theory such as coherent potential approximation (CPA) [24]), the extracted constitutive parameters carry no physical meaning. In Fig. 3(b), we calculate the effective parameters $\varepsilon_{eff}$ ($\mu_{eff}$=1 from the extraction procedure, not shown) using CPA, where even no propagating modes exist near quadrupole resonance 3.9 eV as $\varepsilon_{eff}$ is negative.

### IV. PLASMONIC CLUSTERS

The underlying physics can also be understood from the constitutive relation of macroscopic Maxwell equation: $D_\alpha = E_\alpha + 4\pi P_\alpha - 4\pi \sum_\beta \partial_\beta Q_{\alpha\beta}$ (see Ref. [25], chap. 6). Near the quadrupole resonance, the quadrupole terms cannot be neglected, and therefore the electric permittivity cannot be determined by the electric susceptibility $\chi_e$ alone. The coupling between electric and magnetic dipole resonances is emphasized in metamaterial design. In the optical frequency range, the electric quadrupole terms are actually of the same order of magnitude as magnetic dipole terms, and it is probably prudent to consider the effect of electric quadrupole whenever magnetic dipoles resonances are strong [10], unless one can argue that the electric quadrupole effect is zero due to symmetry or negligible with respect to the magnetic dipole for some particular reason. This point can be illustrated by a prototypical plasmonic lattice system in which the building block comprises of a cluster of metallic nanoparticles. Here we consider only low frequency excitations, and no high multipole ($l \geq 2$) for a single nanoparticle can be excited. In this case every nanoparticle can be treated as a dipole, and we can use coupled dipole equations to describe the system [14]. Let us consider $N$ point dipoles $\mathbf{p}_i$ at $\mathbf{r}_i$, $\{i=1, N\}$. The time dependent current contributed by each dipole is $\mathbf{J}_i = \partial \mathbf{p}_i / \partial t \, \delta(\mathbf{r} - \mathbf{r}_i) = -i\omega \mathbf{p}_i \delta(\mathbf{r} - \mathbf{r}_i)$. By substituting this current to the definition of electric, magnetic dipole and electric quadrupole, and applying the continuity equation $\partial \rho / \partial t + \nabla \cdot \mathbf{J} = 0$ (see Ref. [25], chap. 9), it is straightforward to find

$$\mathbf{P} = \int \mathbf{x} \rho(\mathbf{x}) d^3 x = -\frac{1}{i\omega} \int \mathbf{J} d^3 x \quad , \quad \mathbf{M} = \frac{1}{2c} \int \mathbf{x} \times \mathbf{J} d^3 x \quad , \quad \text{and}$$

$\bar{\mathbf{Q}} = 3\int \mathbf{x}\mathbf{x}\rho(\mathbf{x})d^3x = -\dfrac{3}{i\omega}\int(\mathbf{x}\mathbf{J}+\mathbf{J}\mathbf{x})d^3x$. (Here we use the $\bar{\mathbf{Q}}$ for convenience, the proper definition of quadrupole should be modified to be traceless: $\mathbf{Q} = \bar{\mathbf{Q}} - \tfrac{1}{3}\mathrm{tr}\bar{\mathbf{Q}}$.) The corresponding total electric, magnetic dipole, and quadrupole moments are $\mathbf{P} = \sum \mathbf{p}_i$, $\mathbf{M} = -\dfrac{ik_0}{2}\sum \mathbf{r}_i \times \mathbf{p}_i$, and $\mathbf{Q} = 3\sum(\mathbf{r}_i\mathbf{p}_i + \mathbf{p}_i\mathbf{r}_i)$ respectively. We will show in Appendix I that the EM fields in the far field due to the set of $N$ dipoles is indeed equivalent a radiator with such values of $\mathbf{P}$, $\mathbf{M}$ and $\mathbf{Q}$ as defined by $\mathbf{P} = \sum \mathbf{p}_i$, $\mathbf{M} = -\dfrac{ik_0}{2}\sum \mathbf{r}_i \times \mathbf{p}_i$, and $\mathbf{Q} = 3\sum(\mathbf{r}_i\mathbf{p}_i + \mathbf{p}_i\mathbf{r}_i)$ to the zeroth and first order. We will also consider in Appendix II the more general case where we have a collection of $N$ spheres, which does not have to be small in size. If the particles are small and close together, the $\mathbf{P}$, $\mathbf{Q}$ formula are proved to be correct numerically, while the formula for $\mathbf{M}$ has a very simple analytical form as shown in Appendix II. It is well known that a current loop yields a non-zero magnetic dipole moment, and a vanishing quadrupole moment. However, the quadrupole moment contributed by the current generated by discrete electric dipoles $\mathbf{p}_i$ is usually not zero, and has the same order $L \cdot P$ with the magnetic dipole, where $L$ is the characteristic size of the cluster, and $P$ is the magnitude of electric dipoles. To find a "pure magnetic dipole" ($\mathbf{M} \neq 0$, $\mathbf{Q} = 0$) or a "pure quadrupole" ($\mathbf{Q} \neq 0$, $\mathbf{M} = 0$) in a frequency range, some particular symmetry are required to guarantee one of them to be zero. The behavior of $\mathbf{P}$, $\mathbf{M}$, $\mathbf{Q}$ in a plasmonic cluster could strongly deviate from a single particle since their resonant frequencies rely on the special geometry of the configuration. A useful eigenmode decomposition method has been adopted to investigate dipole clusters [15,26] or one dimensional chain [14]. We choose the configuration of the cluster to be an octahedron, which is shown in the inset of Fig. 4. The radius of sphere is $r_s = 7$ nm, and the spacing between opposite corners is $R = 30$ nm. The octahedron structure has symmetry group $O_h$ which may support negative group velocity in $x$, $y$, $z$ direction simultaneously.

With $\mathbf{P}$, $\mathbf{M}$, $\mathbf{Q}$ defined above, we can classify the eigenstates. Given a particular frequency, we can solve for the eigenvectors that give the values of $\mathbf{p}_i$ on each site, and the $\mathbf{P}$, $\mathbf{M}$, $\mathbf{Q}$ formulae give us the corresponding dipole and quadrupole moments. The octahedral system has many types of multipole resonances. Fig 4(a) shows four different eigenstates below 3.5 eV. The lowest one at 3.32 eV is a magnetic dipole eigenstate, which means only the total magnetic moment **M**≠0, and meanwhile **P**, **Q**=0. The next one at 3.34 eV is a quadrupole state, and the one at 3.44 eV is found to be electric dipole accompanied by another higher multipole with $l \geq 3$. The interaction among dipoles in the octahedron can lower the **P** resonance in frequency from the single particle resonance at about 3.57 eV by 0.12 eV, and it can produce **M** and **Q** resonances. We note that in this particular structure with symmetry group $O_h$, the "pure magnetic dipole" and "pure quadrupole" states can be found and will not mix with each other. However, the **M** and **Q** resonances are rather close in frequency, which reinforces the notion for most plasmonic structures (not just the octahedron) that **M** and **Q** resonances come together. Although we neglect all the dynamic multipoles $l \geq 2$ in such low frequencies for a single sphere, multipole resonances can be generated by geometry of the configuration.

If this plasmonic cluster with octahedron geometry is used as the building block of three dimensional PCs, the corresponding band structure can be classified by the **P**, **M**, **Q** eigenstate defined above. In Fig. 4(b), the band structure of a PC with lattice constant a=75 nm is shown. A negative group velocity band can be observed for quadrupole state near 3.34 eV. Although similar quadrupole bands with negative group velocity have been found for single sphere near 3.9 eV in Sec. II, the more complex building block can lower down the resonance frequency considerably. Many types of building blocks, such as nanorod pairs [4], "fishnet" structures [5], are introduced to realize negative refraction metamaterial in optical frequencies. The negative permeability $\mu < 0$ is usually considered to be associated with a magnetic dipole resonance. Actually for a far field observer, the contribution from electric quadrupoles might be comparable to that from magnetic dipoles, which is demonstrated by simulation [10]. Since in optical frequencies, $\varepsilon$ and $\mu$ are not uniquely defined [8,9], one can describe all electromagnetic responses by dielectric function $\varepsilon(\omega,k)$ with both time and spatial dispersion and set $\mu$ as 1. Or equivalently, especially in the case of metamaterials, people usually prefer a description of $k$-independent effective $\varepsilon$ and $\mu$ to achieve the "double negativity". Our calculation shows the quadrupole itself can render the negative refraction possible even in the absence of electric and magnetic dipole resonance, where the effective medium description by $\varepsilon$ and $\mu$ might not carry useful meaning.

## V. CONCLUSION

In summary, we analytically proved that two branches of quadrupole bands in simple cubic PCs consisting of plasmonic spheres with very small lattice constant, have intrinsic negative group velocities. The mechanism responsible for this new type of negative refraction is analogous to that for *d*-state electrons in transition metals which goes beyond conventional effective medium theory. Since the short range interaction dominates, the QSA is found to be very accurate for quadrupoles as the filling ratio is small. In the typical plasmonic system composed of metallic nanoparticle clusters, the quadrupoles play a significant role because their resonance can be adjusted to lower optical frequencies, and the coupling strength to external EM waves are in the same order of **M**. This can be very important in the design of metamaterials.

## ACKNOWLEDGEMENT

This work was supported by the Central Allocation Grant from the Hong Kong RGC through HKUST3/06C. Computation resources were supported by the Shun Hing Education and Charity Fund. We thank Drs. Xianyu Ao, Jack Ng, and Junjun Xiao for helpful discussions.

## APPENDIX I: EFFECTIVE DIPOLE AND QUADRUPOLE MOMENTS IN THE FAR FIELD

Using Gaussian units, the electric, magnetic dipole and quadrupole magnetic fields in the far field are respectively:

$$\mathbf{H}_P = k_0^2 (\mathbf{n} \times \mathbf{p}) \frac{e^{ik_0 r}}{r}, \tag{I.1}$$

$$\mathbf{H}_M = k_0^2 (\mathbf{n} \times \mathbf{m}) \times \mathbf{n} \frac{e^{ik_0 r}}{r}, \tag{I.2}$$

$$\mathbf{H}_Q = -\frac{ik_0^3}{6} \frac{e^{ik_0 r}}{r} (\mathbf{n} \times \mathbf{Q}(\mathbf{n})). \tag{I.3}$$

Here, **p**, **m**, **Q** represent respectively electric dipole, magnetic diple and electric quadrupole. **n** is the unit vector along the direction **r**. If there are *N* electric dipoles $\mathbf{p}_i$ located at $\mathbf{r}_i$, the magnetic field in the far field is simply the summation of that for every single dipole

$$\mathbf{H} = \sum_{i=1}^{N} \mathbf{H}_i = \sum_{i=1}^{N} k_0^2 (\widehat{\mathbf{r} - \mathbf{r}_i} \times \mathbf{p}_i) \frac{e^{ik_0|\mathbf{r}-\mathbf{r}_i|}}{|\mathbf{r}-\mathbf{r}_i|}, \tag{I.4}$$

where $\widehat{\mathbf{r} - \mathbf{r}_i} = (\mathbf{r} - \mathbf{r}_i)/|\mathbf{r} - \mathbf{r}_i|$ is an unit vector. We expand all the terms including $\mathbf{r}_i$ to the first order:

$$\begin{aligned} |\mathbf{r}-\mathbf{r}_i| &\sim r - \mathbf{n}\cdot\mathbf{r}_i \\ \widehat{\mathbf{r} - \mathbf{r}_i} &= (\mathbf{r}-\mathbf{r}_i)/|\mathbf{r}-\mathbf{r}_i| \sim \mathbf{n} + (\mathbf{n}(\mathbf{n}\cdot\mathbf{r}_i) - \mathbf{r}_i)/r \end{aligned}. \tag{I.5}$$

By substitute (I.5) to (I.4), we find

$$\begin{aligned} \mathbf{H} &= \sum_{i=1}^{N} \mathbf{H}_i = \sum_{i=1}^{N} k_0^2 (\widehat{\mathbf{r} - \mathbf{r}_i} \times \mathbf{p}_i) \frac{e^{ik_0|\mathbf{r}-\mathbf{r}_i|}}{|\mathbf{r}-\mathbf{r}_i|} \\ &= \sum_{i=1}^{N} k_0^2 \{[\mathbf{n} - (\mathbf{n}(\mathbf{n}\cdot\mathbf{r}_i) - \mathbf{r}_i)/r] \times \mathbf{p}_i\} \frac{e^{ik_0(r-\mathbf{n}\cdot\mathbf{r}_i)}}{r - \mathbf{n}\cdot\mathbf{r}_i} \end{aligned}. \tag{I.6}$$

The zero-th order of magnetic field is

$$\mathbf{H}_{(0)} = \sum_{i=1}^{N} k_0^2 (\mathbf{n} \times \mathbf{p}_i) \frac{e^{ik_0 r}}{r} = k_0^2 (\mathbf{n} \times \sum_{i=1}^{N} \mathbf{p}_i) \frac{e^{ik_0 r}}{r}. \tag{I.7}$$

Then the total electric dipole moment for this system can be defined as

$$\mathbf{P} = \sum_{i=1}^{N} \mathbf{p}_i. \tag{I.8}$$

For the first order, we note that the terms proportional to $1/r$ can be ignored in the far field, and only the term $-ik_0 \mathbf{n}\cdot\mathbf{r}_i$ from the phase factor survives in the long wavelength limit $k_0 r_i \ll 1$,

$$\begin{aligned} \mathbf{H} &= \sum_{i=1}^{N} \mathbf{H}_i = \sum_{i=1}^{N} k_0^2 (\mathbf{n}\times\mathbf{p}_i)(-ik_0 \mathbf{n}\cdot\mathbf{r}_i) \frac{e^{ik_0 r}}{r} \\ &= -ik_0^3 \mathbf{n} \times \sum_{i=1}^{N} (\mathbf{n}\cdot\mathbf{r}_i)\mathbf{p}_i \frac{e^{ik_0 r}}{r} \end{aligned} \tag{I.9}$$

The term $(\mathbf{n}\cdot\mathbf{r}_i)\mathbf{p}_i$ in (I.9) can be rewritten as the sum of a symmetric part and an anti-symmetric part:

$$(\mathbf{n}\cdot\mathbf{r}_i)\mathbf{p}_i = \tfrac{1}{2}[(\mathbf{n}\cdot\mathbf{r}_i)\mathbf{p}_i + (\mathbf{n}\cdot\mathbf{p}_i)\mathbf{r}_i] + \tfrac{1}{2}(\mathbf{r}_i \times \mathbf{p}_i) \times \mathbf{n}. \tag{I.10}$$

The anti-symmetric term can be used to define a total magnetic dipole moment

$$\mathbf{M} = -\frac{ik_0}{2} \sum_{i=1}^{N} (\mathbf{r}_i \times \mathbf{p}_i). \tag{I.11}$$

The anti-symmetric part is exactly the same with the result in Ref. [6] by introducing the current $\mathbf{J} = \sum_{i=1}^{N} -i\omega \mathbf{p}_i \delta(\mathbf{r} - \mathbf{r}_i)$.

The symmetric part is

$$\mathbf{H}_{sys} = -ik_0^3 \mathbf{n} \times \sum_{i=1}^{N} \tfrac{1}{2}[(\mathbf{n} \cdot \mathbf{r}_i)\mathbf{p}_i + (\mathbf{n} \cdot \mathbf{p}_i)\mathbf{r}_i] \frac{e^{ik_0 r}}{r} . \qquad (\text{I.12})$$

Comparing this field with the far field radiated from a quadrupole source in Eq. (I.3), we obtain

$$\mathbf{Q}(\mathbf{n}) = \mathbf{Q} \cdot \mathbf{n} = 3\sum_{i=1}^{N} [(\mathbf{n} \cdot \mathbf{r}_i)\mathbf{p}_i + (\mathbf{n} \cdot \mathbf{p}_i)\mathbf{r}_i] . \qquad (\text{I.13})$$

This is consistent with the result in Ref. [25], chap. 9, $\mathbf{Q}(\mathbf{n}) = 3\int \mathbf{r}(\mathbf{n} \cdot \mathbf{r})\rho(\mathbf{r})d^3x = -\frac{3}{i\omega}\int [(\mathbf{n} \cdot \mathbf{r})\mathbf{J} + (\mathbf{n} \cdot \mathbf{J})\mathbf{r}]d^3x$ if the current $\mathbf{J} = \sum_{i=1}^{N} -i\omega \mathbf{p}_i \delta(\mathbf{r} - \mathbf{r}_i)$ is substituted in. We note the trace of $\mathbf{Q}$ contributes a term proportional to $\mathbf{n}$ which does not affect the far field in Eq. (I.12).

So, we see that if there is a collection of dipoles, or a system that can be represented as a collection of discrete dipoles, an observer in the far field will see the a magnetic field (to the lowest order) as if there would be a magnetic dipole moment and an electric quadrupole moment defined by Eq. (I.11) and (I.13) respectively.

## APPENDIX II: VECTOR HARMONICS EXPANSION

The vector spherical harmonics are defined as follows

$$\begin{aligned}
|\mathbf{J}_{lmM}(\mathbf{r})\rangle &= j_l(k_0 r)|\mathbf{x}_{lm}(\hat{\mathbf{r}})\rangle \\
|\mathbf{H}_{lmM}(\mathbf{r})\rangle &= h_l(k_0 r)|\mathbf{x}_{lm}(\hat{\mathbf{r}})\rangle \\
|\mathbf{J}_{lmE}(\mathbf{r})\rangle &= -\frac{i}{k_0}\nabla \times |\mathbf{J}_{lmM}(\mathbf{r})\rangle , \\
|\mathbf{H}_{1mE}(\mathbf{r})\rangle &= -\frac{i}{k_0}\nabla \times |\mathbf{H}_{lmM}(\mathbf{r})\rangle
\end{aligned} \qquad (\text{II.1})$$

where the vector spherical harmonics $\mathbf{x}_{lm}$ are defined as

$$\mathbf{x}_{lm}(\hat{\mathbf{r}}) = -i\mathbf{r} \times \nabla Y_{lm}(\theta, \phi)/\sqrt{l(l+1)} . \qquad (\text{II.2})$$

and $j_l(k_0 r)$, $h_l(k_0 r)$ are respectively spherical Bessel functions and spherical Hankel functions of the first kind. As shown in Ref. [27], the regular and irregular vector spherical solid harmonics can be expanded in terms of a "structure constant"

$$|\mathbf{H}_{lm\sigma}(\mathbf{r}-\mathbf{R})\rangle = \sum_{l'm'\sigma'} G_{lm\sigma;l'm'\sigma'}(\mathbf{R}) |\mathbf{H}_{l'm'\sigma'}(\mathbf{r})\rangle \tag{II.3}$$

where the general form of the structure constants $G_{lm\sigma;l'm'\sigma'}(\mathbf{R})$ can be found in Ref. [27] and references therein. Applying the relations above, we can "shift" the electric dipole **P** at position **R** to electric dipole **P'**, magnetic dipole **M'**, quadrupole **Q'** and higher multipoles at the origin. The structure constants can be simplified by the dipole approximation (namely only electric dipole **P** at position **R**):

$$G_{1mE;l'm'\sigma'}(\mathbf{R}) = \begin{cases} \sum_{\mu} C(1,1,1;m-\mu,\mu) g_{1,m-\mu;l',m'-\mu}(\mathbf{R}) C(l',1,l';m'-\mu,\mu), \\ \quad \text{when}\quad \sigma' = E \\ -\sqrt{\dfrac{2l'+1}{l'+1}} \sum_{\mu} C(1,1,1;m-\mu,\mu) g_{l,m-\mu;l'-1,m'-\mu}(\mathbf{R}) C(l'-1,l',1;m'-\mu,\mu), \\ \quad \text{when}\quad \sigma' = M \end{cases}$$

.

Here $g_{l,m;l',m'}(\mathbf{R})$ are structure factors for the scalar waves, $C$ are the Clebsch-Gordon coefficients between the angular momentum 1 and $l$ which combine the vector nature and spatial dependence of EM waves. The scalar structure factors are given by

$$g_{l,m;l',m'}(\mathbf{R}) = 4\pi \sum_{l''m''} i^{l'+l''-l} C_{lm;l'm';l''m''} j_{l''}(k_0 R) Y_{l''m''}(-\hat{\mathbf{R}}). \tag{II.4}$$

$C_{lm;l'm';l''m''}$ are the Gaunt coefficients which determine the overlap coefficients among three spherical harmonics:

$$C_{lm;l'm';l''m''} = \int d\Omega_k Y^*_{l'm'}(\hat{k}) Y^*_{l''m''}(\hat{k}) Y_{lm}(\hat{k}). \tag{II.5}$$

The electric field for the dipole **P** at position **R** can be expanded as vector spherical harmonics centered at the origin:

$$\mathbf{E}(\mathbf{r}) = \sum_{lm\sigma} b^{\sigma}_{lm} |\mathbf{H}_{lm\sigma}(\mathbf{r})\rangle \tag{II.6}$$

After the rigorous expansion, one should find the relation between the the coefficients of multipoles $l = 1, 2$ to the components of **P'**, **M'**, **Q'** centered at the origin by comparing the electric field with the corresponding electric fileds radiated by **P'**, **M'** and **Q'**. The matrices $\mathbf{T}_{pb}$, $\mathbf{T}_{mb}$, $\mathbf{T}_{Qb}$ can be introduced which satisfy $\mathbf{P'} = \mathbf{T}_{pb} \mathbf{b}^{(E)}$, $\mathbf{M'} = \mathbf{T}_{mb} \mathbf{b}^{(M)}$, $\mathbf{Q'} = \mathbf{T}_{Qb} \mathbf{b}^{(Q)}$. Here, $\mathbf{b}^{(E)}$, $\mathbf{b}^{(M)}$, $\mathbf{b}^{(Q)}$ are the coefficients defined by $\mathbf{b}^{(E)} = (b^{(E)}_{1,-1}, b^{(E)}_{1,0}, b^{(E)}_{1,1})$, $\mathbf{b}^{(M)} = (b^{(M)}_{1,-1}, b^{(M)}_{1,0}, b^{(M)}_{1,1})$, and $\mathbf{b}^{(Q)} = (b^{(E)}_{2,-2}, \cdots, b^{(E)}_{2,2})$. The matrices $\mathbf{T}_{pb}$, $\mathbf{T}_{mb}$ are found to be

$$\mathbf{T}_{pb} = \mathbf{T}_{mb} = \frac{1}{ik_0^3}\sqrt{\frac{3}{16\pi}}\begin{bmatrix} 1 & 0 & -1 \\ -i & 0 & i \\ 0 & \sqrt{2} & 0 \end{bmatrix}, \tag{II.7}$$

while $\mathbf{T}_{Qb}$ is a 5×5 matrix (see Ref. [25], chap. 4) which we are not going to show here. Now we can use $\mathbf{T}_{pb}^{-1}$ to transform electric dipole $\mathbf{P}(\mathbf{r}=\mathbf{R})$ to $\mathbf{b}^{(E)}(\mathbf{r}=\mathbf{R})$, then we can find $\mathbf{b}^{(E,M,Q)}(\mathbf{r}=0)$ at the origin using the known structure constants $G_{lmE;l'm'\sigma'}(\mathbf{R})$. Finally we transform $\mathbf{b}^{(E,M,Q)}(\mathbf{r}=0)$ back to $\mathbf{P'}$, $\mathbf{M'}$, $\mathbf{Q'}$ by $\mathbf{T}_{pb}$, $\mathbf{T}_{mb}$, $\mathbf{T}_{Qb}$ respectively. Numerically we have proved that if the ratio $R/\lambda$ is smaller than 0.1, the total moments $\mathbf{P'}$, $\mathbf{M'}$, $\mathbf{Q'}$ calculated by the far field expansion in Appendix I is correct within the error 10%. Hence, the results shown in Appendix I are good for "small objects", and by small, we mean small compared with the wavelength.

However, the total magnetic moment $\mathbf{M'}$ have a quite simple form since only the $l'=m'=0$ terms contribute to the scalar structure factors $g_{l,m;l',m'}(\mathbf{R})$ given by (II.4). In this case the Gaunt coefficients $C_{lm;l'm';l''m''}$ can be evaluated as

$$C_{lm;00;l''m''} = \int d\Omega_k Y_{00}^*(\hat{k})Y_{l''m''}^*(\hat{k})Y_{lm}(\hat{k}) = \frac{1}{\sqrt{4\pi}}\delta_{ll''}\delta_{mm''},$$

and we have $g_{lm;00}(\mathbf{R}) = \sqrt{4\pi}\, j_l(k_0 R)Y_{lm}(-\hat{\mathbf{R}})$. The structure constants are

$$G_{1mE;1m'M}(\mathbf{R}) = -\sqrt{3\pi}\, C(1,1,1;m-m',m')\, j_1(k_0 R)Y_{1m}(-\hat{\mathbf{R}}),$$

where $C(0,1,1;0,m')=1$ is used. It can be rewritten in a matrix form

$$\mathbf{G}(\mathbf{R}) = \sqrt{3\pi}\, j_1(k_0 R)\begin{bmatrix} Y_{10}(\hat{\mathbf{R}}) & -Y_{1,-1}(\hat{\mathbf{R}}) & 0 \\ Y_{11}(\hat{\mathbf{R}}) & 0 & -Y_{1,-1}(\hat{\mathbf{R}}) \\ 0 & Y_{11}(\hat{\mathbf{R}}) & -Y_{10}(\hat{\mathbf{R}}) \end{bmatrix} \tag{II.8}$$

where we use that $Y_{1m}(-\hat{\mathbf{R}}) = -Y_{1m}(\hat{\mathbf{R}})$. The matrix which "shifts" electric dipole $\mathbf{P}$ at position $\mathbf{R}$ to magnetic dipole $\mathbf{M'}$ at $\mathbf{r}=0$ can be written down as:

$$\mathbf{T}_{mb}\mathbf{G}^T(\mathbf{R})\mathbf{T}_{pb}^{-1} = \sqrt{3\pi}\, j_1(k_0 R)\cdot$$

$$\begin{bmatrix} 1 & 0 & -1 \\ -i & 0 & i \\ 0 & \sqrt{2} & 0 \end{bmatrix}\begin{bmatrix} Y_{10}(\hat{\mathbf{R}}) & Y_{11}(\hat{\mathbf{R}}) & 0 \\ -Y_{1,-1}(\hat{\mathbf{R}}) & 0 & Y_{11}(\hat{\mathbf{R}}) \\ 0 & -Y_{1,-1}(\hat{\mathbf{R}}) & -Y_{10}(\hat{\mathbf{R}}) \end{bmatrix}\begin{bmatrix} \frac{1}{2} & \frac{1}{2} & 0 \\ 0 & 0 & \frac{1}{\sqrt{2}} \\ -\frac{1}{2} & \frac{i}{2} & 0 \end{bmatrix}.$$

After some algebra, we find that

$$\mathbf{T}_{mb}\mathbf{G}^T(\mathbf{R})\mathbf{T}_{pb}^{-1} = \frac{3i}{2} j_1(k_0 R) \begin{bmatrix} 0 & \cos\theta & -\sin\theta\sin\phi \\ -\cos\theta & 0 & \sin\theta\cos\phi \\ \sin\theta\sin\phi & -\sin\theta\cos\phi & 0 \end{bmatrix}. \quad \text{(II.9)}$$

The matrix in Eq. (II.9) may look complicated, but it is exactly the definition of vector product

$$\mathbf{M}' = -\frac{3i}{2} \frac{j_1(k_0 R)}{R} \mathbf{R} \times \mathbf{P}. \quad \text{(II.10)}$$

In the long wavelength limit $k_0 R \ll 1$, $j_1(x) \to \tfrac{1}{3} k_0 R$, Eq. (II.10) becomes $\mathbf{M}' = -\frac{ik_0}{2} \mathbf{R} \times \mathbf{P}$, which is consistent with the result in Appendix I.

### References

1. V. G. Veselago, Sov. Phys. Usp. **10**, 509 (1968).
2. J. B. Pendry, Phys. Rev. Lett. **85**, 3966 (2000).
3. R. A. Shelby, D. R. Smith, and S. Schultz, Science **292**, 77 (2001).
4. V. M. Shalaev, W. S. Cai, U. K. Chettiar, H. K. Yuan, A. K. Sarychev, V. P. Drachev, and A. V. Kildishev, Opt. Lett. **30**, 3356 (2005).
5. S. Zhang, W. Fan, K. J. Malloy, S. R. J. Brueck, N. C. Panoiu and R. M. Osgood, Opt. Express **13**, 4922 (2005).
6. A. Alù, A. Salandrino, and N. Engheta, Opt. Express **14**, 1557 (2006).
7. D. S. Citrin, J. Opt. Soc. Am. B **22**, 1763 (2005).
8. L. D. Landau, E. M. Lifshitz, and L. P. Pitaevskii, Electrodynamics of Continuous Media, (Pergamon Press, New York, 1984), Chap. 9.
9. V. M. Agranovich, Y. R. Shen, R. H. Baughman, and A. A. Zakhidov, Phys. Rev, B **69**, 165112 (2004).
10. D. J. Cho, F. Wang, X. Zhang, and Y. R. Shen, Phys. Rev. B **78**, 121101 (2008).
11. V. L. Moruzzi, J. F. Janak, and A. R. Williams, Calculated Electronic Properties of Metals, (Pergamon Press, New York, 1978), Chap. 4.
12. M. Notomi, Phys. Rev. B **62**, 10696 (2000).
13. C. Luo, S. G. Johnson, J. D. Joannopoulos , and J. B. Pendry JB, Phys. Rev. B

**65**, 201104 (2002).

14. K. H. Fung and C. T. Chan, Opt. Lett. **32**, 973 (2007).

15. K. H. Fung and C. T. Chan, Phys. Rev. B **77**, 205423 (2008).

16. W. Y. Zhang, X.Y. Lei, Z. L. Wang, D. G. Zheng, W.Y. Tam, C. T. Chan, and Ping Sheng, Phys. Rev. Lett. **84**, 2853 (2000).

17. W. H. Weber and G. W. Ford, Phys. Rev. B **70**, 125429 (2004).

18. C. E. Dungey and C. F. Bohren, J. Opt. Soc. Am. A **8**, 81(1991).

19. S. Y. Park and D. Stroud, Phys. Rev. B **69**, 125418 (2004).

20. Y. R. Zhen, K. H. Fung, and C. T. Chan, Phys. Rev. B **78**, 035419 (2008).

21. Y. Wu and Z. Q. Zhang, to be published.

22. D. R. Smith, S. Schultz, P. Markos, and C. M. Soukoulis, Phys. Rev. B, **65**, 195104 (2002).

23. D. R. Smith and J. B. Pendry, J. Opt. Soc. Am. B **23**, 391 (2006).

24. P. Sheng, *Introduction to Wave Scattering, Localization and Mesoscopic Phenomena*, (Springer, New York, 2006), Chap. 4.

25. J. D. Jackson, *Classical Electrodynamics* (Wiley, New York, 1999, 3rd Ed.).

26. V. A. Markel, J. Opt. Soc. Am. B **12**, 1783 (1995).

27. X. D. Wang, X. G. Zhang, Q. L. Yu, and B. N. Harmon, Phys. Rev. B **47**, 4161 (1993).

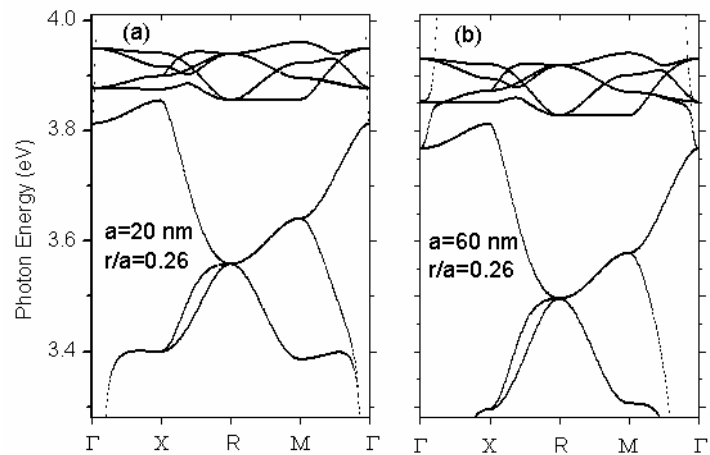

Fig. 1. Band structures of simple cubic PCs consisting of plasmonic spheres

calculated by MST with the ratio $r_s/a = 0.26$ fixed. The lattice constant is 20 and 60 nm in (a) and (b) respectively.

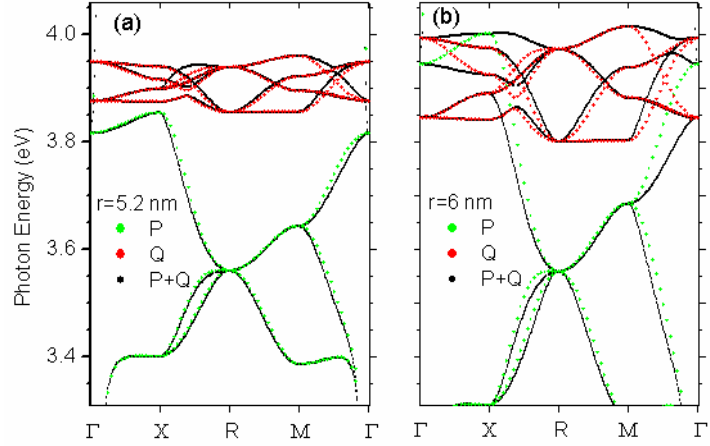

Fig. 2. Band structures calculated by MST with $l =1,2$ (black dots), and $l =1$, $l =2$ separately shown as green and red dots. The sphere radius $r_s$ in (a) and (b) is 5.2 and 6 nm respectively with the lattice constant $a =20$ nm fixed.

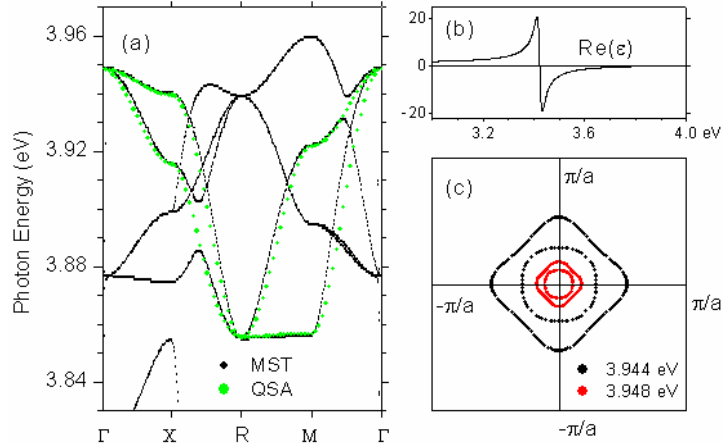

Fig. 3. (a) The dispersion of negative group velocity bands calculated under QSA (green dots) compared with the rigorous MST results (black dots). The lattice constant $a$ is 20 nm and radius $r_s$ is 5.2 nm. (b) The effective permittivity $\varepsilon_{eff}$ ($\mu_{eff}=1$, not shown) calculated by CPA where we use the collision frequency $\gamma =0.02$ eV in Drude model. Quadrupole responses near 3.9 eV are beyond CPA. (c) Equifrequency contours of the two negative bands. The frequency is 3.948 eV (3.944 eV) for red (black) dots. The band with smaller negative slope has square shape even near the Brillouin zone center, while the one with larger negative slope is more isotropic.

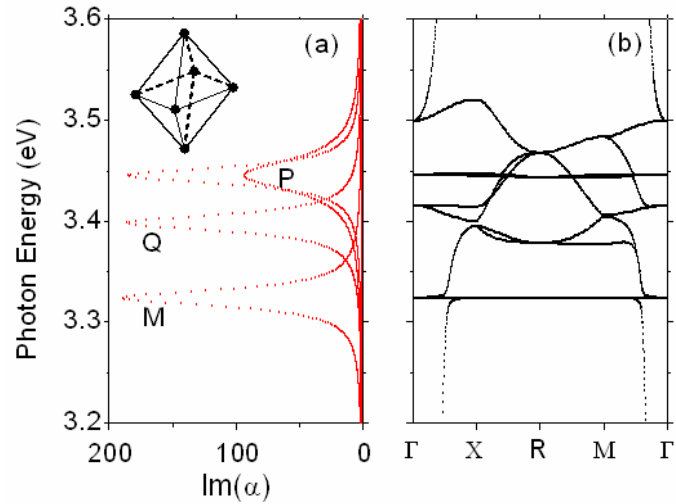

Fig. 4 (a) The imaginary part of the linear response function for an octahedron composed of six sphere with the radius $r_s$ =7 nm, and the spacing between opposite corners is $R$ =30 nm. The collective resonance with the lowest frequency at 3.32 eV is a magnetic dipole. The resonance at 3.4 eV is a quadrupole. And the ones near 3.45 eV are an electric dipole accompanied with another higher multipole. (b) Band structure of a simple cubic PC when the octahedron is used to be the building block in an unit cell. The negative group velocity band can be observed near quadrupole resonance frequency 3.4 eV.